\documentstyle[dina4,11pt,amssym]{article}
\newcommand{\bm}[1]{\mbox{\boldmath$#1$\unboldmath}}

\newcommand{\RP}{ {{\rm \R P}} }
\newcommand{\R}{{\Bbb R}}
\newcommand{\Ls}{{\cal L}}
\newcommand{\Hs}{{\cal H}}
\newcommand{\Ks}{{\cal K}}
\newcommand{\q}{{\bf q}}
\newcommand{\p}{{\bf p}}
\newcommand{\Qb}{{\bf Q}}
\newcommand{\cb}{{\bf c}}
\newcommand{\Fb}{{\bf F}}
\newcommand{\Pb}{{\bf P}}
\newcommand{\G}{{\bf G}}
\newcommand{\X}{{\bf X}}
\newcommand{\x}{{\bf x}}
\newcommand{\J}{{\bf J}}
\newcommand{\r}{{\bf r}}
\newcommand{\Sb}{{\bf S}} 
\newcommand{\A}{{\bf A}}
\newcommand{\B}{{\bf B}}
\newcommand{\Lb}{{\bf L}}
\newcommand{\Id}{\mbox{{\bf 1}}}
\newcommand{\vecII}[2]{\left(\begin{array}{c} #1 \\ #2 \end{array}\right)}
\newcommand{\vecIII}[3]{\left(\begin{array}{c} #1 \\ #2 \\ #3\end{array}\right)}

\newcommand{\matIII}[3]{\left(\begin{array}{ccc}%
 #1 \\ #2 \\ #3 \end{array}\right)}
\newenvironment{mat3}{\left(\begin{array}{ccc}}{\end{array}\right)}
\newenvironment{mat4}{\left(\begin{array}{cccc}}{\end{array}\right)}
\newenvironment{vecn}{\left(\begin{array}{c}}{\end{array}\right)}

\newcommand{\vheta}{\vartheta}
\newcommand{\vhi}{\varphi}

\newcommand{\xib}{{\bm \xi}}
\newcommand{\pib}{{\bm \pi}}
\newcommand{\Tht}{{\bm \Theta}}
\newcommand{\gb}{{\bm \gamma}}
\newcommand{\mac}[1]{ {\check #1} }
\newcommand{\pac}[1]{ {\hat #1} }

\renewcommand{\[}{\begin{equation}}
\renewcommand{\]}{\end{equation}}

\begin{document}

\title{
Inflation of Hamiltonian System: \\
The Spinning Top in Projective Space
}
\author{Holger R. Dullin \\
Institut f\"ur Theoretische Physik \\
Universit\"at Bremen \\
Postfach 330440 \\
28344 Bremen, Germany \vspace{1ex} \\
Email: hdullin@physik.uni-bremen.de
}
\date{April 1996}
\maketitle

\begin{abstract}
We present a method to enlarge the phase space of a canonical
Hamiltonian System in order to remove coordinate singularities
arising from a nontrivial topology of the configuration space.
This ``inflation'' preserves the canonical structure of the system
and generates new constants of motion that realize the constraints.
As a first illustrative example the spherical pendulum is inflated
by embedding the sphere $S^2$ in the three dimensional Euclidean space.
The main application which motivated this work is the derivation
of a canonical singularity free Hamiltonian for the general 
spinning top. The configuration space $SO(3)$ is diffeomorphic to 
the real projective space $\RP^3$ which is embedded in 
four dimensions using homogenous coordinates. 
The procedure can be generalized to $SO(n)$.
\end{abstract}

\section{Introduction}

One of the pillars of classical mechanics is the process 
of deriving equations of motion via the Lagrangian $L$. It starts with
some ``generalized coordinates'' on the configuration space $Q$
as described, e.g., in \cite{Whittaker37,Gold80,Arnold78}. 
The Euler-Lagrange equations of motion are then obtained from 
the variational principle that $\int \Ls \,{\rm d}t$ be extremal on 
solutions of the mechanical system.
For most configuration spaces $Q$ there are no global singularity free 
coordinate systems which could be used as generalized coordinates in 
the Lagrangian. In the classical examples with compact configuration space
(e.g., spherical pendulum, spinning top, geodesic flow on the ellipsoid,
see e.g.\ \cite{Arnold78,Jacobi1866,Whittaker37})
global coordinates with singularities are used. This is well suited, e.g., 
in order to apply the method of separation of variables to solve the 
equations of motion.
For integrable systems the coordinate singularity is typically
only encountered by orbits with special values of the constants of
motion. In the nonintegrable case the constants of motion are absent,
while the coordinate singularity is still present. Moreover, it is
unforseeable which orbits will encounter the singularity.
Especially for the numerical integration these coordinate systems
are therefore not advisable.

There are two possible cures. Since $Q$ is a manifold there exist
local coordinates everywhere, such that we obtain Lagrangians in
every chart, supplemented by the transition maps between different
charts. This approach is certainly the most general, but it is 
neither very elegant nor very simple.
The second approach starts with an embedding of $Q$, $\dim Q = n$,
as a submanifold of $\R^{n+k}$. In order to fulfill the constraints
that define the submanifold $Q$, Lagrange multipliers are introduced, and
the equations of motion are obtained by standard procedures, 
see e.g.~\cite{Gold80}.
Kozlov \cite{Kozlov96} introduced the notion of ``excessive coordinates'',
i.e.\ the description of a mechanical system with more 
coordinates than degrees of freedom. The passage from a Lagrangian
with multipliers to a Hamiltonian in excessive coordinates 
is described in \cite{Kozlov96}. Our approach is different in that
we start with the Hamiltonian in (singular) generalized coordinates
and employ a transformation to a Hamiltonian in (nonsingular) 
excessive coordinates, which has all the constraints as additional
constants of motion. 
This ``inflation transformation'' is described in the first part
and illustrated for the case of the spherical pendulum.
In the second part we show how to apply this method to the spinning top.
After this work, which first appeared in my thesis \cite{Dullin94b}, 
was completed, I learned that Kozlov also obtained 
the same Hamiltonian for the spinning top \cite{Kozlov96}.
Nevertheless, the method at hand gives a quite different derivation of
these equations, which also generalizes to $SO(n)$.

\section{Inflation of Hamiltonian Systems}

Consider a Hamiltonian with $n$ degrees of freedom $\Hs(\q,\p)$.
We introduce $n+k$ new coordinates $\Qb$ by
\begin{eqnarray}
\label{eqn:KanoTraf}	\q & = & \Fb(\Qb) \\
\nonumber		\cb_q & = & \tilde \Fb(\Qb).
\end{eqnarray}
We think of the old configuration space, parametrized by $n$ coordinates $\q$,
as extended by $k$ additional coordinates $\cb_q$ all of which do {\em not} 
show up in the Hamiltonian, i.e.\ we introduce $k$ cyclic variables.
The question now is, how to introduce corresponding new canonical momenta 
$\Pb$ in such a way that the desired geometric constraints $\tilde \Fb(\Qb)$
are constants of motion of the dynamics of a new Hamiltonian depending
on $(\Qb,\Pb)$. Note that we must allow the transformation 
to have singularities at the points where the old coordinates $\q$ have 
the coordinate singularities that we are going to remove.

This is achieved by taking (\ref{eqn:KanoTraf}) as a 
point transformation from $\Qb$ to the trivially extended coordinates
$(\q,\cb_q)$. Using the generating function
\[
	S = \Fb(\Qb) \p + \tilde \Fb(\Qb) \cb_p
\]
we recover (\ref{eqn:KanoTraf}) for the coordinates by construction, while
the momenta are given by 
\[
 \Pb   =  
   \left( \frac{ \partial(\Fb,\tilde \Fb) }{\partial \Qb}\right)^t
	 \vecII{\p}{\cb_p}.
\]
The inverse yields the desired transformation to the new momenta:
\[
\label{eqn:Impis}
\vecII{\p}{\cb_p}  = 
	{\left( \frac{ \partial(\Fb,\tilde \Fb) }{\partial \Qb}\right)^t}^{-1} \Pb \\
        =:  \vecII{ \G(\Qb,\Pb) }{ \tilde \G(\Qb,\Pb) }.
\]
With this trick (the rest are standard canonical transformations) we can show:
\goodbreak
{\em
The dynamics of the Hamiltonian System 
$\Ks(\Qb,\Pb) = \Hs( \Fb(\Qb), \G(\Qb,\Pb) )$
is equivalent to the dynamics of the Hamiltonian System $\Hs(\q,\p)$, i.e.:
\begin{enumerate}
\item  The solutions $(\Qb(t),\Pb(t))$ of the inflated system 
are mapped onto solutions of the original system $(\q(t),\p(t))$ by
$(\Fb,\G)$.
\item  $\tilde \Fb(\Qb)$ and $\tilde \G(\Qb,\Pb)$ 
are constants of motion of the inflated system.
\item  The canonical 1-forms are equal on solutions of the systems:
$\p \, d\q = \Pb \, d\Qb$.
\end{enumerate}
}

These statements are clear, since the transformation is (in an extended sense)
a canonical transformation. Once we have the enlarged system, we can
transform it back to the old one via a standard canonical transformation,
with the unusual result that $k$ coordinates {\em and} $k$ momenta
become cyclic. This is possible because the new Hamiltonian is
singular, i.e. the Hessian of $\Hs$ with respect to $\p$ is degenerate.
Because of this we can not pass back to a Lagrangian in these variables.
The above three statements are comprised in the transformation of the
Poisson bracket.
With the notation $\X := (\Qb,\Pb)^t$ and
$\x := (\q,\cb_q,\p,\cb_p)^t$ we obtain, for any function $f$,
\begin{eqnarray}
\{ f , \Ks \}_X & = & (\nabla_X f)^t \J \nabla_X \Ks  \\
	& = & \nabla_x f^t \frac{\partial x}{\partial X} \J 
		(\frac{\partial x}{\partial X})^t \nabla_x \Hs  \\
	& = & \nabla_x f^t \J \nabla_x \Hs = \{ f, \Hs \}_{q,p}.
\end{eqnarray}
Since $\Hs$ does not depend on $\cb_q$, $\cb_p$, in the last row we
obtain the bracket on the original space.

\subsection*{The Spherical Pendulum}

As a toy example we consider the spherical pendulum described by the 
Hamiltonian
\[ \label{eqn:sphendel}
	\Hs = \frac{1}{2}\left(
	p_\vheta^2 + \frac{p_\vhi^2}{\sin^2\vheta}\right) - g \cos\vheta,
\]
where the singular coordinates $(\vheta,\vhi)$ are used to
parametrize the configuration space $S^2$. 
We take the obvious embedding of $S^2$ in $\R^3$ given by $x^2+y^2+z^2=1$,
i.e.\ we take the coordinates $\r=(x,y,z)^t$ of the real $\R^3$ in which 
the pendulum moves, as excessive coordinates:
\begin{equation}
	\vecIII{\vhi}{\vheta}{c_q} = \vecIII
			{\arctan y/x}
			{\arccos z/r}
			{x^2 + y^2 + z^2} =: \vecIII{F_1}{F_2}{\tilde F}.
\end{equation}
With the notation $\rho^2 = x^2 + y^2$ we obtain for the Jacobian
\[
\frac{\partial (\Fb,\tilde F) }{\partial \r} = \matIII
	{ -\frac{y}{\rho^2}   & \frac{x}{\rho^2}    &  0}
	{ \frac{xz}{\rho r^2} & \frac{yz}{\rho r^2} & -\frac{\rho}{r^2} }
	{ 2x                  & 2y                  & 2z },
\]
such that the momenta are given by
\[
\vecIII{p_\vhi}{p_\vheta}{c_p} = \matIII
	{ -y      & x       & 0     }
	{ xz/\rho & yz/\rho & -\rho }
	{ x/2r^2   & y/2r^2   & z/2r^2 } \vecIII{p_x}{p_y}{p_z}.
\]
Carrying out the transformation, one obtains a singularity free
Hamiltonian for the spherical pendulum in the symmetric form
\[ \label{eqn:sphepe}
	\Ks = \frac{1}{2}\left(
	(xp_y-yp_x)^2 + (xp_z-zp_x)^2 + (yp_z-zp_y)^2\right) - \frac{z}{r}.
\]
A short computation shows that in addition to the old constant of motion
$p_\vhi = l_z = xp_y-yp_x$, 
there are geometric constants of motion
$r^2$ and $\p \r/2r^2$ representing the constraint.
By an appropriate choice of initial conditions we can achieve
$r^2 = 1$ and $\p \r = 0$, so that the geometric constraints are
the sphere and the orientation of its tangent plane in $\R^3$, 
as must be the case.

Neither this coordinate system nor Hamiltonian (\ref{eqn:sphepe}) are new
(according to Klein and Sommerfeld \cite{KS10} it was first 
used by Hermite). As stated in the introduction, our focus is on the
transformation procedure, which allows the application to more
complicated systems like the spinning top.

\section{The Spinning Top in Projective Space}

As usual we take ``spinning top'' to be a short hand for the
motion of a rigid body around a fixed point -- with or without
a potential. The configuration space of the spinning top is
$SO(3)$. This manifold is parametrized (again with singularities)
by Euler angles. To remove these singularities we choose the
Cayley parametrization of $SO(3)$, which immediately generalizes
to $Q=SO(m)$: Consider the antisymmetric 
$m \times m$ matrices $\A$ with $\dim Q = n = m(m-1)/2$ independent entries
$\xi_1,\xi_2,\ldots,\xi_n$. 
Now introduce an additional coordinate $\xi_0$ and map the $(n+1)$-tuple 
$\xib := (\xi_0,\xi_1,...,\xi_n)$ 
onto a matrix $\B \in SO(m)$ via Cayley's map $C$
\begin{equation} \label{eqn:Cayley}
	C : \xib \mapsto \B = (\xi_0 \Id - \A)^{-1} (\xi_0 \Id + \A).
\end{equation}
Obviously every line of $\xib$ is mapped to the same $\B$, since
$C(\xib) = C(\lambda \xib)$ for any $\lambda \in \R \setminus \{0\}$.
Thus the 
preimage $C^{-1}(\B)$ is a line through the origin, so we can endow the
$\xib$ with the structure of the projective space $\RP^n$. 
The components of $\xib$ are then interpreted as homogenous 
coordinates, and $C$ as a diffeomorphism
\[
	C : \RP^n \rightarrow SO(m).
\]
Thus we have $k=1$ for the global homogeneous coordinates, where
the constraint is obtained from choosing a representative for
every projective line on $S^n$, by restricting $\xib$ to unit length.
Antipodal points of $S^n$ have the same image under C.
If we identify these points we obtain a model of $SO(m)$ as
\[
	S^n / \{\pm 1\} \simeq \RP^n \simeq SO(m).
\]
For $m=n=3$ our equations live on $S^3\simeq SU(2)$, 
the universal covering space of $\RP^3\simeq SO(3)$. In this case
the coordinates $\xib$ are usually called Euler parameters.
For the rotation matrix we obtain
\[
\label{eqn:so3C}
\B^t = \frac{1}{\xib^2}\begin{mat3}
 2\xi_0^2+2\xi_1^2-\xib^2& 2\xi_1\xi_2 -2\xi_0\xi_3& 2\xi_0\xi_2+2\xi_1\xi_3 \\
 2\xi_1\xi_2 +2\xi_0\xi_3& 2\xi_0^2+2\xi_2^2-\xib^2&-2\xi_0\xi_1+2\xi_2\xi_3\\
-2\xi_0\xi_2 +2\xi_1\xi_3& 2\xi_0\xi_1+2\xi_2\xi_3 & 2\xi_0^2+2\xi_3^2-\xib^2
\end{mat3}.
\]
The vector $(\xi_1,\xi_2,\xi_3)^t$ is an eigenvector of $\B$ with eigenvalue 1.
Thus it is the fixed axis of rotation described by $\B$. The angle of 
rotation $\alpha$ is given by $\cos (\alpha/2) = \xi_0/2$.
The description of rotations by the 4-vectors $\xib$ is directly related to the
Pauli spin matrices respectively the algebra of quaternions, 
see e.g.\ \cite{KS10,Gold80,Whittaker37,Ebinghaus88}.
We only remark that both choices of ``spin'', $+\xib$ and $-\xib$, by
construction give the same $\B$.

Comparing the matrix $\B^t$ with an $SO(3)$ matrix given in Euler angles
in the convention used in \cite{Gold80}, we 
can read off the inflation transformation to obtain a Hamiltonian with
$\xib$ as coordinates for the configuration space and
constraint $\xi^2=1$:
\begin{eqnarray}
\vhi & = & \arctan \frac{\xi_0\xi_2 - \xi_1\xi_3}{\xi_0\xi_1 + \xi_2\xi_3} \\
\vheta & =&\arccos \frac{\xi_0^2-\xi_1^2-\xi_2^2+\xi_3^2}{\xi^2} \\
\psi & = & \arctan \frac{\xi_0\xi_2 + \xi_1\xi_3}{-\xi_0\xi_1 + \xi_2\xi_3} \\
c_q  & = & \sqrt{\xi^2}
\end{eqnarray}
The old momenta given by the new ones $\pib$ are
\begin{eqnarray}
p_\vhi & = & 
		\frac{1}{2} (\xi_3\pi_0 - \xi_2\pi_1 + \xi_1\pi_2 -\xi_0\pi_3) \\
p_\vheta &=& 
    \frac{1}{2}\sqrt{\frac{\xi_0^2+\xi_3^2}{\xi_1^2+\xi_2^2}}
	(\xi_1\pi_1-\xi_2\pi_2)-
    \frac{1}{2}\sqrt{\frac{\xi_1^2+\xi_2^2}{\xi_0^2+\xi_3^2}}
	(\xi_0\pi_0-\xi_3\pi_3) \\
p_\psi & = & 
		\frac{1}{2} (\xi_3\pi_0 + \xi_2\pi_1 + \xi_1\pi_2 + \xi_0\pi_3)\\
c_p & = & \xib^t \pib/\xi.
\end{eqnarray}
Inserting this into the original Hamiltonian in Euler angles 
(see, e.g., \cite{Gold80,DJR94})
the coordinate singularities are removed and we obtain the very
symmetric Hamiltonian
\begin{eqnarray}
\nonumber
\Hs & = &
	\frac{(\pi_0\xi_1-\pi_1\xi_0+\pi_2\xi_3-\pi_3\xi_2)^2}{8\Theta_1}+
	\frac{(\pi_0\xi_2-\pi_1\xi_3-\pi_2\xi_0+\pi_3\xi_1)^2}{8\Theta_2}+\\
& & 
	\frac{(\pi_0\xi_3+\pi_1\xi_2-\pi_2\xi_1-\pi_3\xi_0)^2}{8\Theta_3}
	+ \frac{1}{\xi^2}\{ (2\xi_0\xi_2+2\xi_1\xi_3)S_1 + \\
\nonumber & & 
	(-2\xi_0\xi_1 + 2\xi_2\xi_3)S_2 + (2\xi_0^2+2\xi_3^2-\xib^2)S_3 \},
\end{eqnarray}
where we have used a linear potential acting on the center of mass 
$\Sb$, and the moments of inertia $\Tht$.
Introducing the notation
\begin{eqnarray}
\xib = \begin{vecn} \xi_0 \\ \xi_1 \\ \xi_2 \\ \xi_3 \end{vecn}
	\quad \Leftrightarrow \quad
	\pac\xib, \mac\xib = \begin{mat4}
			\xi_1 & -\xi_0   & \pm\xi_3 & \mp\xi_2 \\
			\xi_2 & \mp\xi_3 & -\xi_0    & \pm\xi_1 \\
			\xi_3 & \pm\xi_2 & \mp\xi_1  &   -\xi_0 \\
		\end{mat4},
\end{eqnarray}
(the upper signs belong to $\pac\xib$ and the lower to $\mac\xib$)
and $\mac\xib_z$ for the third row of $\mac\xib$ we can
write the Hamiltonian most compactly as
\begin{equation} \label{eqn:HamPro}
	\Hs = \frac{1}{8} \pib^t \pac\xib^t \Tht^{-1} \pac\xib \pib +
		\frac{1}{\xi^2} \pac\xib \mac\xib_z^t \cdot \Sb.
\end{equation}
The constants of motion induced by the constraint are $\xib^2$ and 
$\xib^t \pib$, which can be fixed to $1$ and $0$ respectively,
by an appropriate
choice of initial conditions. These two constants of motion are 
related to the identities $\mac\xib \xib = \pac\xib = 0$ and
$\pac\xib\pib + \pac\pib\xib = 0$.
The angular momentum in body coordinates
is $\Lb = \pac\xib \pib/2$, and in space fixed coordinates $\mac \xib \pib/2$.
The reason for these symmetric expressions is the relation 
$\mac\xib = \B \pac\xib$. Using $\pac\xib\pac\xib^t = \xi^2 \Id$ this 
gives a factorization of $\B = \mac\xib\pac\xib^t/\xi^2$.
The total angular momentum is given by $4l^2=\xi^2\pi^2 - (\xib^t\pib)^2$,
and with the above choice of values for the geometric constraints 
it just gives $l^2 = \pi^2/4$.

Note that the Hamiltonian (and the constants of motion) are invariant 
under a change of ``spin'', i.e.\ under
the transformation $(\xib,\pib) \rightarrow (-\xib,-\pib)$, which is 
induced by the transformation to the antipodal point on $S^3$.
Our coordinates are from $T^\ast S^3$, but the dynamics takes place in 
$T^\ast SO(3)$. E.g.\ for a periodic orbit which is non contractable
in $SO(3)$ we will find twice its period in $S^3$. Most notably this
applies to all the relative equilibria.

If we introduce $\gb = \pac\xib \mac\xib^t_z/\xi^2$ for the
coordinates of the space fixed unit vector in $z$-direction
we can write the equations of motion as
\[
	\dot \xib = \frac{1}{2} \pac\xib^t \Tht^{-1} \Lb \qquad
	\dot \pib = \frac{1}{2} \pac\pib^t \Tht^{-1} \Lb - 
\left( \frac{\partial \gb}{\partial \xib} \right)^t \Sb.
\]
Finally we remark that since $l_z$ is also a constant of motion, we can
introduce new coordinates $(\gb, \Lb)$ as given above, and perform a
reduction to the standard Euler-Poisson-equations on $T^\ast S^2$.

\section{Discussion}

We have shown how to describe a Hamiltonian system in a configuration
space with higher dimension than the number of degrees of freedom.
The resulting description with excessive coordinates preserves the
Hamiltonian structure. This description is always useful when the
configuration space has a nontrivial topology such that there
do not exist global coordinates. Our main motivation was to 
obtain singularity free {\em and} Hamiltonian equations of motion, 
which are needed, e.g.\ for the numerical calculation of 
actions in integrable cases as described in \cite{DW94,DJR94}.
The use of Euler parameters, which are used as a global coordinate system,
has a long history. Weierstra\ss{} \cite{Weierstrass1897} 
and Klein \& Sommerfeld
\cite{KS10} used them to obtain explicit solutions for integrable cases.
Also Whittaker \cite{Whittaker37} and Goldstein \cite{Gold80} give
an introduction to the description of rotations with Euler parameters.
However they don't give a Hamiltonian in these variables.
Related descriptions are also given by Kirchgraber and Stiefel 
\cite{Kirch78} for the use in perturbation theory, and in
\cite{Evans77} and \cite{Sonnenschein85} for numerical integration.
Let us again remark that Kozlov \cite{Kozlov96} gives the same
equations, but with a different derivation. 
Our approach using Cayleys map directly generalizes to $SO(n)$.

\section{Acknowledgement}

The author would like to thank Peter H. Richter and Andreas Wittek
for inspiring discussions. This work was supported by the
Deutsche Forschungsgemeinschaft.

\bibliographystyle{plain}

\end{document}